\documentclass[conference]{sig}
\usepackage{color, xcolor, enumerate, graphicx, url, multirow, xspace, hyperref}
\usepackage[font=bf, skip=0pt]{caption}

\newcommand{\BfPara}[1]{{\noindent\bf#1.}\xspace}
\newcommand{\etal}{{\em et al.}\xspace}

\usepackage{listings}
\usepackage{fancyvrb}
\usepackage{verbatim}
\usepackage{seqsplit}
\usepackage{color,soul}
\usepackage[framemethod=TikZ]{mdframed}
\usepackage{mathtools}
\usepackage{amsmath}
\usepackage{balance}
\usepackage{algorithm}
\usepackage{algpseudocode}
\makeatletter
\renewcommand{\ALG@name}{Scoring Procedure}
\makeatother

\DeclareMathOperator*{\argmin}{arg\,min}

\usepackage{xspace}
\newcommand{\qoi}{{\sf QoI}\xspace}

\begin{document}

\title{Beyond Free Riding: Quality of Indicators for Assessing  Participation in Information Sharing for Threat Intelligence\thanks{Approved for Public Release; Distribution Unlimited: 88ABW-2017-0416, Dated 01 Feb 2017. This work was done while O. Al-Ibrahim was visiting the University at Buffalo.}}
\author{Omar Al-Ibrahim \\ University at Buffalo \and Aziz Mohaisen\\University at Buffalo \and
Charles~Kamhoua \\ Air Force Research Lab \and
        Kevin Kwiat \\ Air Force Research Lab \and
        Laurent Njilla\\  Air Force Research Lab
}

\maketitle



\begin{abstract}
Threat intelligence sharing has become a growing concept, whereby entities can exchange patterns of threats with each other, in the form of indicators, to a community of trust for threat analysis and incident response. However, sharing threat-related information have posed various risks to an organization that pertains to its security, privacy, and competitiveness. Given the coinciding benefits and risks of threat information sharing, some entities have adopted an elusive behavior of ``free-riding'' so that they can acquire the benefits of sharing without contributing much to the community. So far, understanding the effectiveness of sharing has been viewed from the perspective of the amount of information exchanged as opposed to its quality. In this paper, we introduce the notion of quality of indicators (\qoi) for the assessment of the level of contribution by participants in information sharing for threat intelligence. We exemplify this notion through various metrics, including correctness, relevance, utility, and uniqueness of indicators. In order to realize the notion of \qoi, we conducted an empirical study and taken a benchmark approach to define quality metrics, then we obtained a reference dataset and utilized tools from the machine learning literature for quality assessment. We compared these results against a model that only considers the volume of information as a metric for contribution, and unveiled various interesting observations, including the ability to spot low quality contributions that are synonym to free riding in threat information sharing.
\end{abstract}

\section{Introduction}\label{sec:intro}
Today, the Internet connects millions of users, networks, and network collections worldwide, where the Internet's security and stability are quite important to the global economy and well-being of the human race. However, challenged by various forms of cyber-attacks, ensuring the security of the Internet and combatting the various attacks requires proper reconnaissance that prelude countermeasure actions. The information security and threat landscape has grown significantly, making it difficult for a single defender to defend against all of these attacks alone. As such information sharing for threat intelligence, a paradigm in which threat indicators are shared in a community of trust to facilitate defenses, has been on the rise~\cite{Mohaisen2016Rethink}. 

In threat intelligence sharing, participants exchange patterns of threats with each others, in a form of threat indicators or signals. Participants are defined over a community of trust, and collaborate towards a common goal or mission; to understand and respond to emerging threats~\cite{ToshSKKM15}. 
For such intelligence sharing to happen, standards for representation, exchange, and consumption of indicators are proposed in the literature~\cite{Martin2008,barnum2012,kampanakis2014}. Communities of trust are established, and systems and initiatives for sharing are built. For such initiatives to work, participants need to contribute information in those systems to be consumed by other  community members. However, sharing threat-related information have posed various risks to organization, which pertain to security, privacy, and competitiveness. Given the coinciding benefits and risks of threat information sharing, some community members have adopted an elusive behavior of ``free-riding''~\cite{feldman2005} so that they can achieve utility of the sharing paradigms without contributing much to the community.

So far, understanding the effectiveness of sharing has been viewed from the point of view of whether participants contribute or not using volume-based notions of contributions. Thus, a community member who does not contribute a volume of data (indicators) is a free-riding community member~\cite{tanaka2005vulnerability}. The state-of-the art on the problem did not include other metrics beyond simple measures of volume-based contribution, particularly metrics that capture and assess the Quality of Indicators (\qoi) as a mean of understanding contribution in the information sharing paradigms. 

\subsection{Quality of Indicators}
We believe that the nature of the information sharing as a concept and its application to threat intelligence both make quality a very relevant notion, and call for further investigation into the notion's definition and quantification in various settings. A well-defined measure of \qoi could provide a better way of capturing contribution in general, and distinguishing community members who contribute useful data. Furthermore, threat intelligence systems present distinct challenges and opportunities to counter the problem of free-riding and other abusive behaviors once quality is defined. On the one hand, these systems generally lack the enforcement of a central authority, and therefore entities in these systems share information at their own will in a (somewhat) peer-to-peer fashion. As a result, it is necessary to envision a set of notions and mechanisms that characterize contribution in general, and are capable of capturing free-riding, while being implemented in a distributed manner and used by each community member.
An ideal measure of \qoi should be robust to distinguish between the various members based on their contribution, rather than a predefined notion of trust. With the possible speciality of community members, and the varying usefulness of indicators shared based on the context in which they are used, a major challenge is to assign context-dependent quality markers for indicators.

\subsection{The Simple Contribution Measures}
To the best of our knowledge, the problem of free-riding in information sharing for threat intelligence, while sparsely mentioned in other work~\cite{ToshSKKM15,ToshSMKK15}, is not treated properly in the literature. Thus, this work is the first of its type to be dedicated to the problem by identifying \qoi as a new metric of contribution to capture free-riding in information sharing for threat intelligence.  

Given the aforementioned challenges, one simple measure of contribution in information sharing systems weigh the volume of indicators contributed by various community members. However, for many reasons, such measure is insufficient, as described earlier. Therefore, it is important to understand the quality of shared information as a form of participation. Without a high quality of shared information, we cannot achieve actionable intelligence that is effective in combating cyber threats. Unfortunately, this issue is not well understood in the literature, and requires further exploration by identifying the meaning of quality, and basic methods and tools for assessing them are lacking. 

\subsection{Features of Quality of Indicators}
In~\cite{Mohaisen2016Rethink}, Mohaisen \etal explored the potential correlation between \qoi and privacy. However, privacy is not the only factor that affects \qoi. One feature of quality is the \emph{correctness} of an indicator; a meaningful annotation and label of the indicator that is true and accurate. A second possible feature of quality is the \emph{relevance} of the indicator to the community members; because of the targeted nature of modern cyberthreats, information that is shared has to be contextual to the domain. A third plausible feature of quality of an indicator is its \emph{utility}; informally, some indicators are more indicative than others about cyber-attacks, and therefore it is critical that participants in the threat intelligence community share information that capture prominent features of cyber-threats. Finally, the \emph{uniqueness} of an indicator is another assessor of quality, which is defined as a measure of (dis)similarity with previously seen indicators. This property ensures that participants deliver indicators that are not duplicates or redundant, and provide additional threat information to other community members.

Besides these features, indicators are often time-sensitive, making temporal features very important when evaluating \qoi. A timely indicator such as a source of an attack could be used to defend against an emerging attack, unlike a stale indicator that could be (potentially) used for postmortem analysis. As mentioned above, there is also potential correlation between \qoi and privacy. Privacy can affect \qoi (although perhaps negatively when privacy of indicator is ensured). We elaborate on this quality metric in this work, and show its quantification through data-driven analysis.

While each of the aforementioned measures can be used as a separate feature of quality, we envision that a single indicator could have multiple of those features. As such, we also assess \qoi with respect to these metrics in the form of a weighted (continuous) score. Our method for evaluating \qoi is based on exploiting a fine-grained historical records as benchmark for assessing the contributions of community members. We illustrate the concept through a concrete evaluation of a real dataset from various security vendors of antivirus scans and their results of labeling malware samples as seen in the VirusTotal service (\url{https://virustotal.com/}).  

\subsection{Contributions} The main contributions of this paper are multifold. First, we identify the need for \qoi to capture contributions by community members in information sharing paradigm. \qoi captures a wide spectrum of behaviors, from altruistic behavior, where a community member contributes a lot of (high quality) indicators to free-riding, where a community member contributes less, or contribute a lot of low quality indicators. Second, we develop and formulate various metrics that are robust to capture the notion of quality. Third, we experimentally demonstrate those measures and metrics, and show their robustness, and how they differ in identifying contributor's behavior (particularly free-riding) from the simple volume-based measure of contribution.

\subsection{Organization} The organization of this paper is as follows. In Section~\ref{sec:overview}, we provide an overview on cyber-threat intelligence and the risks of information sharing. In Section~\ref{sec:methodology}, we provide an overview of our quality of indicator(QoI) assessment methodology. In Section~\ref{sec:qoi}, we describe the processes involved in our QoI-based assessment. In Section~\ref{sec:results}, we present the results of our benchmark experiment, afterwards we discuss related work in Section~\ref{sec:related} and finally we conclude in Section~\ref{sec:conclusion}.

\section{Overview and Preliminaries}\label{sec:overview}
We first provide an overview of cyber-threat intelligence systems, then introduce to unique problem with information sharing in these systems which demand quality measures.

\subsection{The Threat Landscape} 
The Internet today connects hundreds of millions of users worldwide, and is operated by service providers who connect businesses, education institutes, and government agencies, collectively forming a global village. In the recent years, the Internet has been challenged by various forms of cyber attacks, ranging from endpoint malware attacks~\cite{MekkyMZ15} to massive network disruptions and instabilities~\cite{WangMCC15}. 

At the endpoint side, malware is capable of penetrating a perimeter's security in many enterprise systems, exfiltrating sensitive data from such systems, and causing great damage to both private and public sector networks~\cite{Mohaisen15}. At the larger scale, multiple endpoint infections by malware are more powerful, and pose a greater risk, seen often in systemized large-scale botnets~\cite{WangMCC15b}. Botnets, defined as collections of networks of infected hosts are the basic fabric for the operation of many cybercriminal activities. Botnets rely on principled designs, where bots (infected hosts in a botnet) execute commands on behalf of their herder (botmaster), utilizing command and control (C\&C) infrastructure~\cite{ThomasM14b}.  

Botnets today are used for a variety of cybercriminal activities, including spam, massive denial-of-service (DDoS) attacks, and data exfiltration, among many others. Botnets represent a major component of the cybercrime ecosystem, with the rise of botnet-as-a-service. Today, hackers utilize network reconnaissance to probe targets for vulnerabilities and craft custom payloads to gain control over their infrastructure by spreading malware in propagation efforts. 

\subsection{The Need for Threat Intelligence}
Defending against the threat vectors of malware and botnets is a challenging task, which resulted in a rich body of literature. The  literature on defending against malware and botnets looks into identifying ``signals'', ``indicators'', or simply ``features'' that could be useful in identifying endpoint systems, malware, and botnets. For malware, for example, such features could include static strings in that piece of malware, dynamic artifacts that the malware generates when executed in the wild, or external context information associated with the binary binary of the malware (such as the author's information, operating system, etc.). For botnets, the C\&C infrastructure may include domain names and Internet Protocol (IP) addresses, and knowing such information can be very helpful in identifying a botnet. For example, botnets tend to use Domain Generation Algorithms (DGAs)~\cite{yadav2010detecting}, which result in random domain strings with high entropy, and being able to identify those domain names is key to detection of botnets. Furthermore, being able to distinguish between various DGAs is key to attribution of threat to a certain botnet family.  An effective cyber defense would rely on a good visibility into many of those features. 

Combatting cyber threats and attacks requires intelligence gathering that prelude countermeasure actions, as seen in the above examples. To this end, cyberthreat intelligence has become a growing concept. Today, organizations in the public and private sector, government and industry, have established tools seeking first-hand knowledge about new cyber-attacks and malware threats. This includes the ability to recognize and act upon indicators of attack and compromise scenarios, essentially putting the pieces together for analysis about attack methods and processes using static and dynamic analysis and profiling techniques, open source, social media, and dark web intelligence. 

\subsection{Threat Intelligence Sharing} 
The need for information sharing for threat intelligence is necessitated by both economical and technical realities. Being able to identify all the types of indicators and features useful for characterizing, identifying, and defending against all types of threats, while desirable, is infeasible from both technical and economical standpoints. With new technologies such as cloud, mobile computing, social networks, and the Internet of Things (IoTs), and the the persistence of adversaries through cybercrime and advanced persistent threats (APTs) have also brought several challenges. Therefore, it is reasonable to say that no single player in this ecosystem is capable of addressing all security issues alone.

For this reason, sharing information of threat intelligence among vendors and government entities has emerged as a plausible technique for efficiently and effectively defending against new and emerging threats. With threat intelligence sharing, operational experience is communicated to other parties in an ecosystem to enable them to effectively defend against current attacks, and to improve their defense posture by preventing such attacks from happening utilizing such actionable intelligence. 

To enable information sharing, organizations need to agree on standardizing threat information. This requires defining the content fields, encoding, and exchange format of the information relevant to a particular threat or incident, along with a pre-defined protocol to communicate the criticality of such information.  Various standards for information sharing have been proposed~\cite{Martin2008, barnum2012, kampanakis2014} to automate and structure the exchange of threat information with a community of trust.  

Today, standards are used in the exchange of indicators of software, hardware, and network artifacts, and are intended for operationalizing those indicators in many applications, including security operations related to malware characterization, vulnerability analysis, remediation, platform hardening, and incident-response~\cite{Mohaisen2016Rethink}. 

\subsection{Risks of Sharing} While threat information sharing brings many benefits to the sharing community members, it may incur security risks about participants, their operational contexts, and security posture.  Not only that, the same information, once exposed to an adversary, may be used to test their applicability on other target systems, who may lag behind in security updates or miss out on patching vulnerabilities. Therefore, the adversary will be able to utilize such information for attacking other unpatched systems.

The risk of sharing may go beyond fingerprinting systems to leaking personal identifiable information about individuals.  
Various types of sharing standards are proven to leak personal identifiable information (PII) that may contain names, email addresses, and other types of sensitive data~\cite{friedewald2007privacy}. For example, privacy violations in sharing standards may occur in the form of a document which contains contact information for the constituent responsible for an incident report. This type of information may become personally identifiable in the case when the contact information of a particular individual are used.

Participants in a threat intelligence sharing community may interact with one another with various degrees of collaboration and competition, which may affect the way they share~\cite{ToshSMKK15,ToshSKKM15}. Because of that, many companies and organizations today are reluctant in sharing  firsthand intelligence, and mostly gather and ingest information from neighboring sources that are less significant. 

\subsection{Formulation of the Free-Riding Problem} Given the triad of security, privacy, and risks associated with threat intelligence sharing, some members might be joining communities of sharing for the purpose of benefiting from the platform without offering valuable information themselves, hence, the term, ``free-riding'' is coined to refer to the behavior of such users who act to maximize their own utility at the expense of the welfare of the community. 

This problem is not new, and is manifested in other distributed settings, most notably Peer-to-Peer (P2P) systems. In P2P systems, cooperation is required for the operation the system. However, cooperation may incur significant communication and computational overhead, thus users may refuse to contribute their fair share of resources. At the same time they may utilize the system by consuming the resources of other peers. Researchers demonstrated the impact of free-riders in P2P systems, such as BitTorent~\cite{locher2006}, and observed a significant increase in download times for high contributing nodes in the presence of few low contributing ones.

\subsection{How Quality of Indicators Help}  
So far, and to the best of our knowledge, understanding the effectiveness of sharing has been viewed from the point of view of whether participants contribute or not (thus the literal meaning of free-riding).  This form of contribution is perceived as a volume-based contribution, since the level of contribution by any participant is evaluated directly by the amount of information communicated to the community regardless of its nature, whether it is used by community members or not. Given the large amount of unprocessed threat-related events, which are generated by automation tools, such as security information and event management (SIEM) technologies, and the fact that in many sharing systems today, the contribution level amounts to the volume of data, actors may find it more convenient to submit raw, unprocessed, or unused events as indicators of threat to avoid the investment on resources for cleaning, contextualizing, operationalizing, and filtering such information.  For this reason, it is important to consider the quality of shared information as the basis for evaluating the level of participation, because a simple and coarse measure of participation is insufficient. In order to overcome these obstacles, in this paper we propose QoI as a quantifiable and measurable metrics and provide a methods for quality assessment.

\section{QoI Assessment Methodology}\label{sec:methodology}

Assessing \qoi is a nontrivial task. However, as other quality notions (e.g., quality of services, quality of experience, etc.), \qoi requires defining metrics, methodology for assessing such metrics, and methods for validation of the proposed metrics based on sound assumptions, to find out what capabilities they provide, including addressing free-riding. 

Our approach in developing \qoi metrics is intuitive, and uses several sounds assumptions driven from the context in which indicators are used. In particular, \qoi metrics include correctness of an indicator with respect to a label feature, the relevance of the indicator to a consuming community member, the utility of the indicator, and its uniqueness. More (informal) details are provided in section~\ref{sec:qoimetrics}.

Our approach for assessment of \qoi uses a reference ``golden'' dataset as ground truth. Based on whether the indicator provided by a community member is in the reference dataset or not, the assessor proceeds by either matching the metric attribute of the indicator to that of the golden dataset. If the indicator is not in the golden set, and assuming that the golden dataset has ``similar indicators'', a machine learning algorithm is used for predicting the attribute of the \qoi metric, and compare it to the one provided by community member. For an arbitrary number of indicators provided by the community member, a score is then established for that community member based on the normalized weighted sum of \qoi values across all indicators. 

In this section, we describe the quality metrics and the details of our methodology including a system architecture for assessment, processes followed and the data-flow.

\subsection{QoI Metrics}\label{sec:qoimetrics}
In the following, we identified four metrics to be used for assessment of quality: correctness, relevance, utility, and uniqueness, as described below.

\subsubsection{Correctness} 
For a given reference dataset, the correctness metric of \qoi captures whether attributes of an indicator (e.g., label used for attribution, severity score used for risk assessment, etc.) are consistent with the assessor's reference. For that, and using the labeling of a malware sample as an example of an attribute for an indicator, we  compute the correctness score as the fraction of samples that match the anticipated labels. Specifically, this is computed as the aggregate binary score of the correct samples normalized by the size of the sample set of indicators.

\subsubsection{Relevance} 

Informally, the relevance metric of a \qoi measures the extent to which an indicator submitted by a community member to the community is contextual and of interest to the rest of the community. As such, in defining and assessing the relevance of an indicator, we use a reference weight assignment to the class labels giving higher weight to labels of greater interest to a particularly community member (assessor) and lower weights to less desirable labels. The relevance score is then computed as the average weighted sum for all sample indicators in the set.

\subsubsection{Utility} 
The utility is similar to the relevance of an indicator, although at a finer-granularity than an indicator. As such, we view the utility of an indicator as the average weighted sum of all of its feature components. This is, we assign a different weight to each feature of the indicator to leverage features that are a better candidate input to prediction of threats. While the weighting of the features of an indicator could be realized using one of many ways, we suggest the information gain as a measure of weighting features. For example, using a similar notion, the weights of the feature components can be computed using various statistical models for sensitivity analysis, including the Principal Component Analysis (PCA) technique~\cite{jackson2005}).

\subsubsection{Uniqueness} 
The uniqueness of an indicator is a measure of the (dis)\-similarity of the indicator in comparison with other submitted indicators by contributors in the community. A vector distance (e.g., using the Mahalanobis distance~\cite{mahalanobis1936}, which captures the difference between a point $\vec{x}$ and a distribution of points $X=\{\vec{x_1}, \vec{x_2}, \dots, \vec{x_n}\}$ with a mean $\vec{\mu}$) is computed to determine the degree of uniqueness. We also define a threshold on the minimum distance between the feature vector of an indicator to other indicators, and use that threshold to tell whether an indicator is unique or not.

\subsection{System Architecture \& Design}
Having elaborated on the informal definition of \qoi metrics, we now move to discuss the \qoi system architecture, first as a strawman  highlighting the main concept of assessing \qoi, and then as a fully functioning system that addresses various issues in the strawman design.

\begin{figure}[h]
\centering
\includegraphics[width=0.5\textwidth]{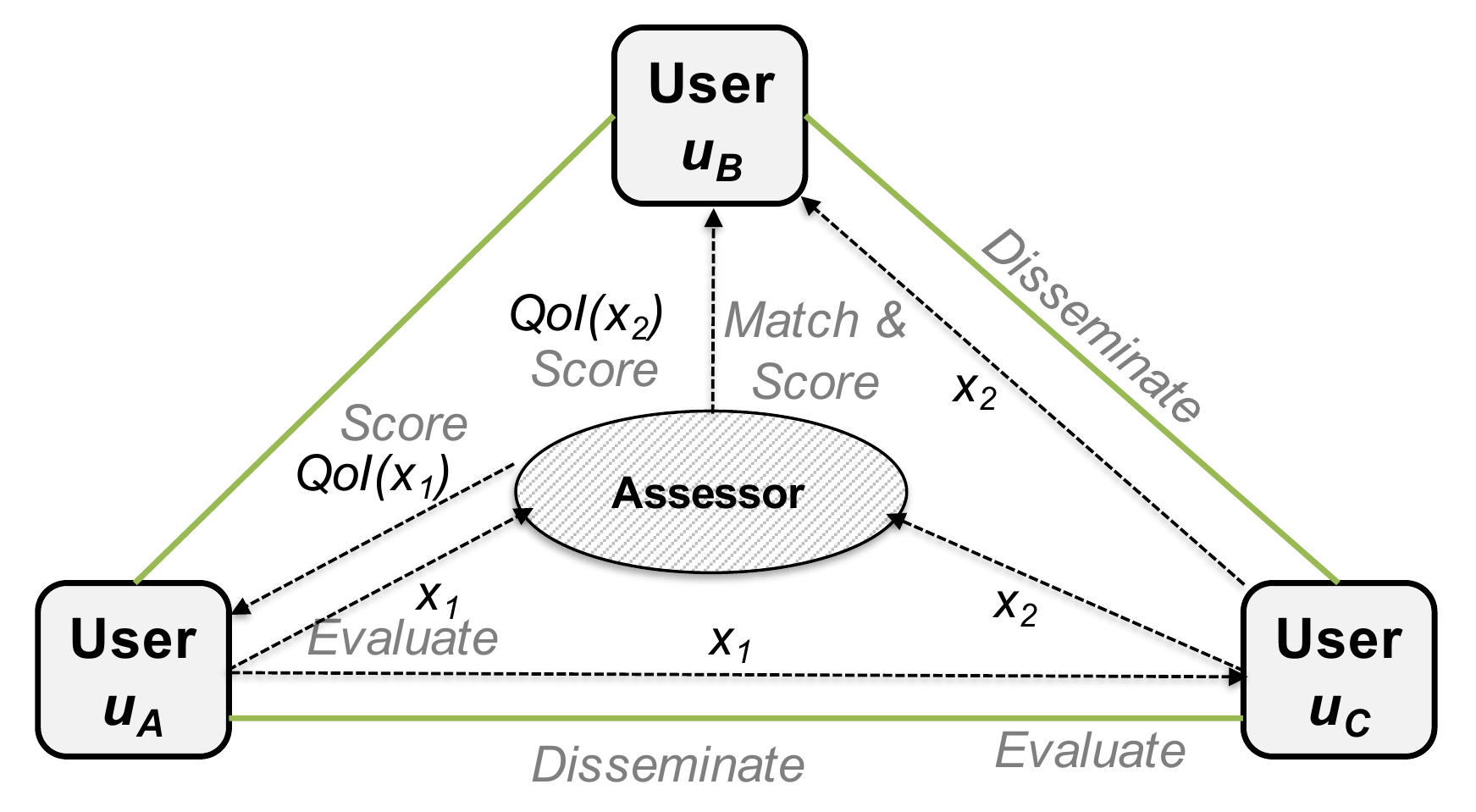}
\caption{A strawman design for an architecture to assess \qoi using a centralized assessor.}
\label{fig:arch}
\end{figure}

Our system for assessing \qoi operates for a set of distributed nodes in a community of trust. Those nodes are are logically connected with each other in a P2P fashion, as shown in Figure~\ref{fig:arch}. As such, each of these nodes would participate in the sharing and consumption of threat indicators provided by other peers, which is achieved within a community of trust that is separated from other communities in the sharing ecosystem. Before nodes can accept and operationalize (process) these threat indicators, they need to evaluate their quality by asking a special node, an assessor, which has sufficient information to perform such function, for a rating (scoring) of the indicator. In response, the assessor assigns a quality score for the indicator based on a ground-truth available to the assessor, and using a reference dataset the assessor has access to. 

In this architecture, we assume that the messages sent between the peers in the system are authentic and tamper-evident, using existing threat exchange protocols that provide end-to-end security guarantees. One issue that the above architecture may suffer from is the amount of trust each community member has to put in the assessor, and the validity of his scoring of indicators. We address this issue by assuming that the architecture can support more than one assessor node, and these nodes may coordinate among each each other.  We leave exploring the spectrum of the number of assessors and the rationale for various numbers as a feature work. However, as a feature, a consumer of a score by the assessors (where the assessors' number is greater than 2) could perform a majority voting to improve the robustness of the scoring to address intentional bias (dishonesty of assessors, if any), and unintentional bias (due to issues with the underlying ground truth dataset). On the other hand, any and every node in the system could act as an assessor, if the golden reference dataset is available to them. 

While the latter assumption of the availability of the data to each node in the system is very implausible, a milder assumption for the operation of the strawman design above is the coverage of data: the system assumes the reference dataset has sufficient information about every possible indicators presented by the various community members. However, based on a prior work assessing coverage of indicators~\cite{Mohaisen2014}, no single community member (antivirus scanner) in the case of malware detection and labeling has a 100\% coverage or accuracy. Based on the same study, and for a malware family such as Zeus~\cite{MohaisenA13}, it takes 6 and 18 community members to provide close to perfect coverage of detection and correctness of labeling, respectively. Such numbers are close to 10\% and 30\%, respectively, of the entire set of community members with antivirus scans in the VirusTotal dataset. This in particular calls for a more ``intelligent'' process for the assessment of \qoi, using not only explicitly provided labels, but also using learned labels from features of indicators utilizing advanced machine learning techniques.

\subsection{System Setup and Steps}\label{sec:procedure}

At a high-level, our ideal system for assessment has the following specific procedures for system setup:
\begin{enumerate}
\item \emph{Defining quality metrics and scoring procedures.}  Quality metrics are used as a measurement criteria to ensure that community members who participate in information sharing provide threat indicators that are valuable to other members, while scoring procedures are methods that specify how these metrics are used to generate a quality score.
\item \emph{Defining annotations for threat and quality labeling.} Annotations can either be labels that indicate the type of threat or they can be labels for identifying the quality (severity, timeliness, etc.) level or quality type of an indicator. Utilizing these annotations, a weight value is assigned to each quality label, and a scoring method is utilized to convert the quality labels to a numeric aggregate score for the indicator. 
\item \emph{Building the reference dataset.} The reference dataset will be used to evaluate \qoi for a sample of indicators submitted by a sample provider. To build the initial reference dataset, data that is collected through security operations (e.g., monitoring, profiling, analyses, etc.) is vetted for their validity and applicability to the domain, perhaps using often expensive by necessary manual vetting~\cite{Mohaisen2014,MohaisenA14a}.
\item \emph{Defining extrapolation procedures and training the classifier.} Extrapolation procedures enable a quality assessor to predict the label of an indicator using its feature set and classifier model. The classifier is trained using a supervised learning process extracted from the reference dataset. This reference dataset is collected for the purpose of initializing the system. 
\end{enumerate}

After the initial setup of the system, the sample indicator is assessed for its quality and a quality score is computed. The following describes the steps of the assessment. 1) Obtain a set of sample indicators where each sample is composed of a tuple (label, vector) that consists of a label and a vector of features. 2) For each sample, extract the feature vector and feed the data as test input to the trained classifier which predicts its label. 3) Compare between the predicted label and the label provided by the sample. Indicate whether the two labels match, and record the comparison result as a quality annotation. 4) Compute the confidence level and include other quality annotations for the indicator using the labeling rubric. 5) Use scoring procedures and quality labels to compute a quality score for the indicator.


\subsection{QoI Assessment Process Operation}

\begin{figure}[t]
\centering
\includegraphics[width=0.5\textwidth]{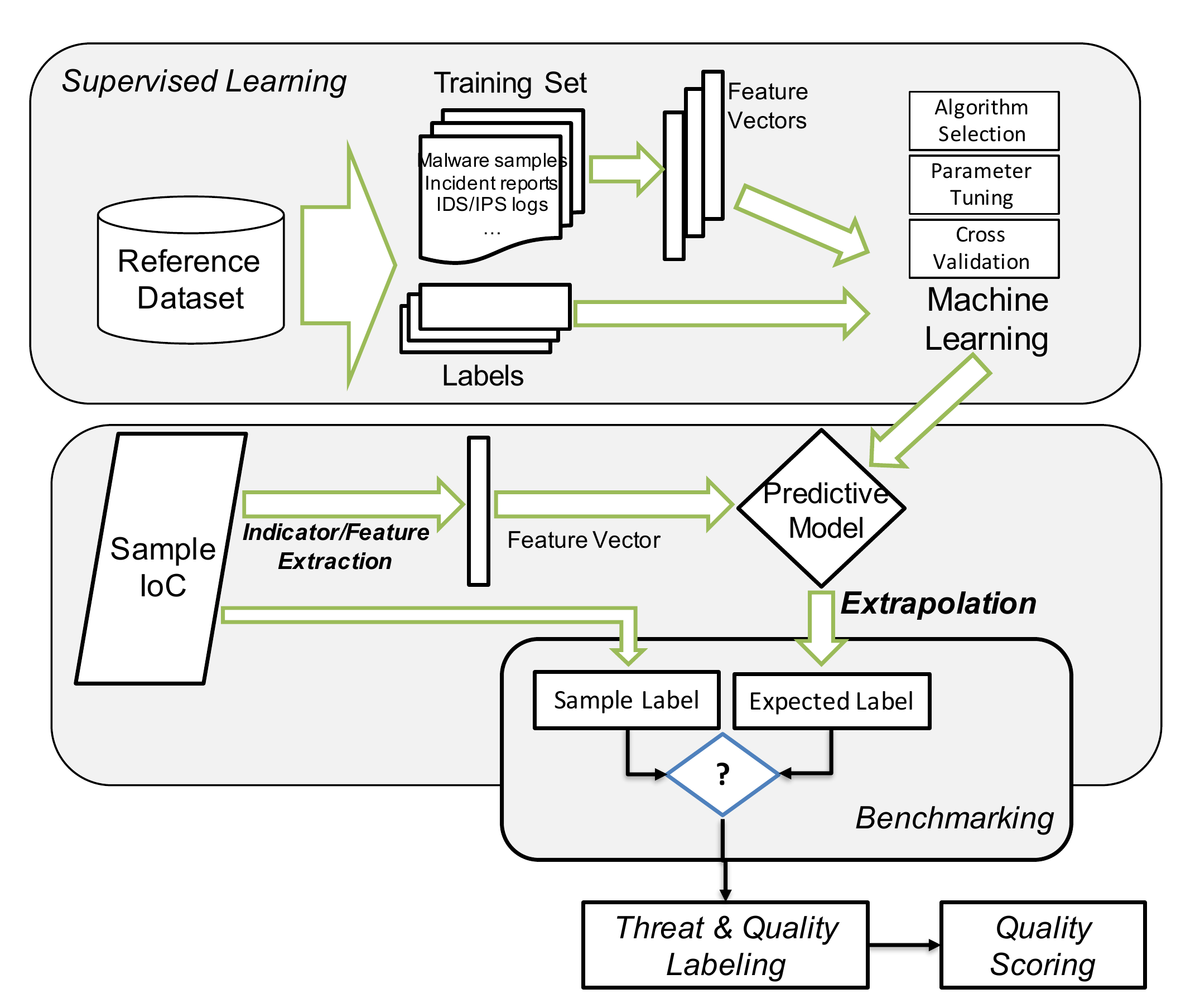}
\caption{The \qoi assessment process, incorporating a reference model established using the training procedure of a machine learning algorithm and predicted label of an indicator provided by a community member. For correctness of indicators as the target metric, a high quality is achieved when the predicted label matches the label provided by the community member (sample IoC label).}
\label{fig:method}
\end{figure}

The illustration for the complete process and dataflow of the \qoi assessment method embraced in our design is depicted in Figure~\ref{fig:method}. As can be seen in the figure, the setup of the system is achieved through the use of a supervised learning process over the reference dataset, rather than the direct matching of explicit labels of indicators in the dataset. 

In this assessment system, we assume a reference labeled (training) dataset that contains a comprehensive library of artifacts, such as malware samples, incident reports, and logs, and that has been collected through typical operational intelligence gathering procedures. Ways for obtaining such labels falls out of the scope of this work, and industry's best practices, as described in~\cite{Mohaisen2014} could be used. Upon ingesting those artifacts in our system, they are  converted into a set of training samples with their corresponding features. Each sample is a pair that consists of a feature vector as an input object for the machine learning algorithm and the corresponding threat label as the desired output value for the object class. To reach such end goal of predicting a label correctly, the build of our trained model encompass multiple components, namely a feature selection procedure, a machine learning algorithm selection procedure (e.g., SVM, logistic regression, random forest, etc.) and the corresponding parameters (e.g., procedure for regularization and linearization in case of SVM and LR, respectively), and cross validation procedures (e.g., fold size, validation strategy, etc.).

Upon building the machine learning model, establishing a confidence in its performance through typical evaluation metrics (e.g., low false positive and false negative, and high true positive and true negative), we then use the built model as a predictor for previously unseen threat indicators. In such operational setting, given a sample indicator provided by a community member, and before ingesting it the community member would pass it to the assessor for further evaluation and scoring. The assessor then extracts a feature set corresponding to the indicator using a standard form, and converts it into a feature vector. The assessor then uses the previously built model as a predictor, and assigns a label (e.g., using a multiclass SVM, the assessor can indicate the label closest in the training set to that of the newly observed indicator). The assessor then decides the quality of the indicator by taking both the predicted and the self-provided label by the community member into account.  The quality scoring engine then uses the individual scores of the various indicators provided by each community member to assess their actual contribution, and detect free-riders.

We note that the ``intelligent'' system above addresses various issues in the strawman system. First, rather than requiring the actual indicators to be present in the ground-truth dataset, this technique requires only the availability of a sufficient number of indicators of the same label. Second, with such flexibility in defining the ground-truth through a learning and model-building process, the number of community members that can act as assessors greatly increase. Finally, the even when the labels of indicators are not provided by the community member, e.g., artifacts provided to the community are not operationalized, this technique makes use of those indicators through other measures of \qoi, such as relevance, or utility, or uniqueness, which do not require a label to be provided by the contributing community member.

\section{QoI Assessment Procedures}\label{sec:qoi}

As discussed in section \ref{sec:procedure}, the \qoi assessment process is composed of a series of steps in order to initialize and operate our system for both assessment of individual indicators and scoring of community members as a whole. Specifically, these steps begin with collecting the reference dataset and building the prediction model, then extrapolating, benchmarking, and computing a quality score for a given indicator. In the following, we outline in more details each of those processes and procedures.

\subsection{Reference Dataset and Learning}

After identifying metrics for defining quality, as exemplified in section~\ref{sec:qoimetrics}, we demonstrated the use of \qoi for the assessment of the contribution level of participants.  As mentioned before, our methods for computing the \qoi involve multiple processes. In order to initialize the system the reference dataset is used to build prediction model through supervised techniques of machine learning.

Specifically, this involves submission of sample artifacts from multiple sources, and the premise is to utilize the notion of quality as opposed to a simple view of contribution based on the volume of data (i.e. the number of samples). In order to evaluate the \qoi  provided by community members, a reference dataset is used as a resource of ground truth. 

While proposing methods for obtaining ground truth falls out of the scope of this work, we only use an example in this work to bootstrap the evaluation. In particular, in this work we demonstrate evaluating the quality of malware labels by AV vendor using VirusTotal as a reference dataset with samples that are manually vetted by one community member~\cite{Mohaisen2014} (in such setting VirusTotal could be loosely defined as the community of trust). In short, VirusTotal is a multi-engine AV scanner that accepts submissions by users and scans the samples with those engines. The results from VirusTotal provide many useful artifacts and annotations, including the labeling of a sample by the various engines of AV scanners, as well as other behavioral and static features of malware samples. Though there might be some inconsistencies in the final labeling between the results of VirusTotal and across vendors, the premise is that the tool can be trusted for samples that have been submitted multiple times over sufficient periods of time, particularly since AV vendors update their results with VirusTotal whenever acquiring a new signature for a previously unknown sample.


Features provided in the ground truth are particularly useful in a learning a model for their label prediction. To identify the family to which a malware sample belongs, security vendors (AV scanners, community members) usually gather various characteristics and features of the sample using static analysis, dynamic analysis, and memory forensics. For static analysis, artifacts like file name, size, hashes, magic literals, compression artifacts, date, source, author, file type and portal executable (PE) header, sections, imports, import hash, among others, are used. For dynamic analysis, file system, user memory, registry, and network artifacts and features are collected. For memory forensics, memory byte patterns are captured to create a signature.

\subsection {Extrapolation and Benchmarking}

After building the prediction function by training classifier with the reference dataset, the next step is to  assess \qoi through extrapolation from the prediction function results. The remaining question becomes how the reference set are used to assess and extrapolate the values and quality of indicators.  In order to answer this question, we elaborate on a particular machine learning techniques, the semi-supervised learning and its application to the problem at hand. 

\BfPara{Classifier model}
While our system described in the previous section uses multiple off-the-shelf algorithms, we highlight the operation of \qoi using a classifier model based on the nearest centroid classifier (ncc), specifically we adopt a variant called the linear discriminant analysis~\cite{izenman2013} to map threat indicators to their respective labels. In this model, each label is characterized by its vector of average feature values (i.e. class centroid). A new sample indicator is evaluated by computing the scaled distance between the features of the sample and each class centroid, and then the sample is assigned to the class to which it is nearest.

To build the classifier, we obtain $r$ samples for training from the reference dataset. This dataset is built such that there are $r_{i}$ training samples per class, with $d$ features per sample. For each training sample $y$, we observe a label $\ell \in \Lambda$ and a sample vector $\vec{y}$. For simplicity we refer to the classes labels by their indices $i=1,2, \ldots, \lambda$. Note that each $\vec{y}$ is a vector of length $m$. We assume that samples labeled by $i$ are distributed as $\mathcal{N}(\mu_i , \Sigma)$, the multivariate normal distribution with mean vector $\mu_i$ and standard deviation matrix $\Sigma$. We denote by $\mathcal{L}(x, \mu_i, \Sigma)$, the corresponding probability density function. Finally, let $\pi_i$ be the prior probability that an unknown sample comes from class labeled by $i$. 

Bayes' Theorem states that the probability that an observed sample $x$ comes from class $i$ is proportional to the product of the class density and prior probability:
\begin{equation}
P(Z = i | X = x)  \propto \mathcal{L}(x, \mu_i, \Sigma) \times \pi_{i}
\end{equation}
where $P(Z = i | X = x)$ is the posterior probability that sample $x$ comes from class $i$. The classifier assigns the sample to the class with the largest posterior probability to minimize the misclassification error. This can be written as a rule:
\begin{equation}\label{eq:2}
\hat{z}(x)  = \arg \min_i \{ (x- \mu_i)^T \Sigma^{-1} (x- \mu_i) - 2\log(\pi_i) \}. 
\end{equation}
Therefore, a sample is assigned to the nearest class and the distance is computed using the LDA metric: 
$||x - \mu_i||^2_{\Sigma} - 2\log(\pi_i)$, where $||x-\mu||^2 = (x-\mu)^T \Sigma^{-1}(x-\mu)$ is the square of the Mahalanobis distance between $x$ and $\mu$.

\BfPara{Misclassification rate}
A misclassfication occurs when an indicator is assigned to an incorrect label. The probability of making a classification error $P(\epsilon)$ is:
\begin{equation}
P(\epsilon) = \sum_{j=1}^{ \lambda } [ P(\hat{Z} \neq j | Z = j) \times \pi_j ].
\end{equation}
The misclassification rate using the LDA rule can be derived from (\ref{eq:2}). In particular, we can calculate the misclassification rate of the nearest-centroid using:
\begin{equation}
P(\epsilon) = \sum_{j=1}^{ \lambda }[ 1- \phi(\underset{i \neq j}{\min}\left\{ \frac{\|\mu_j - \mu_i\|_{\Sigma}^2+2\log(\frac{\pi_j}{\pi_i})}{2\|\mu_j-\mu_i\|_{\Sigma}} \right\}  ) ] \times \pi_{j}, 
\end{equation}
\noindent where $\phi$ is the cumulative distribution function {(cdf)} of the standard normal distribution. Note that this assumes that the sample data are normally distributed as stated by the model. The equation above can be interpreted as a measure of the collective distance between all of the class centroids taking into account class prior probabilities. In general, the misclassification rate is small when the centroids are far apart and increases otherwise.

\subsection{Labeling and Quality Scoring}

We used the ncc to predict labels for observed indicators and compared the results with the sample labels. This enables us to compute a score on the correctness and quality of the feature set for the indicators. In the following we formalize the steps to compute a score for the samples based on the quality metrics described earlier.

Denote by $n$ the number of users in the system.
Each user $u_i$ provides a set of samples 
$X_i=  \{  (\vec{x}_{i1}, l_{i1}), (\vec{x}_{i2}, l_{i2}), (\vec{x}_{i3}, l_{i3}),$ $\ldots ,(\vec{x}_{ik}, l_{ik}) \}$ with feature vector $\vec{x}_{ij}$ and sample label $l_{ij} \in \Lambda$ for $i=1,2,  \dots, n$ and $j=1,2, \ldots,k$.

\subsubsection{Correctness}
As described earlier, the reference dataset is used as the benchmark for determining the correct label for an arbitrary sample. Each sample consists of a feature vector and an associated label. In procedure~\ref{alg:correct} we outline the algorithm used for computing the correctness as a \qoi metric.

\begin{algorithm}[t]
\caption{Correctness of $X_i$ (C)}
\label{alg:correct}
\begin{algorithmic}[1]
\State Obtain reference dataset $Y=\{  (\vec{y}_{1}, l_{1}), (\vec{y}_{2}, l_{2}), (\vec{y}_{3}, l_{3}),$ $\ldots ,(\vec{y}_{r}, l_{r}) \}$ where $l_i \in \Lambda$ for $i=1,2, \ldots , r$. 

\State Evaluate $X_i$ by applying the ncc method as follows: 
\begin{enumerate}[a]
\item \emph{Training}: Compute reference indicators $\vec{\mu}_1, \vec{\mu}_2, \ldots , \vec{\mu}_{|\Lambda|}$ for class labels in $\Lambda$, as per-class centroids $\vec{\mu}_\ell=1/|Y_t|\sum_{(\vec{y}_i,l_i) \in Y_\ell} \vec{y}_i$, where $Y_\ell$ is a subset of $Y$ belonging to the class label $\ell \in \Lambda$.

\item \emph{Prediction}: For every sample $\vec{x}_{ij}$, apply the classifier function to compute the label, $$l^\prime= \argmin_{l \in \Lambda} || \vec{\mu}_l - \vec{x}||$$
\end{enumerate}
\State For every sample, compute the sample score $s_c(\vec{x}_{ij})$ as:
\begin{align}
s_c(\vec{x}_{ij})=
 \begin{cases}
  1 & l^\prime_j = l_{ij} \\
  0 & \text{otherwise}  \nonumber
\end{cases}
\end{align}

\State Compute the correctness score (C) of $X_i$ by taking the average sum:
$C(X_i)= \frac{1}{k} \sum_{j=1}^{k} s(\vec{x}_{ij})$

\end{algorithmic}
\end{algorithm}

As shown above, we first build a classifier by utilizing the reference dataset $Y$ as the training set and forming a prediction on the label of $\vec{x}$ to obtain $l^\prime$. Then, the assigned label of $\vec{x}$ is compared against the predicted label $l^\prime$ and a positive score is given if labels match. The correctness is computed as the average sum of scores for all samples in  $X_i$.

\subsubsection{Relevance}
The steps for computing the relevance variable of a set of indicators are shown in Scoring Procedure~\ref{alg:relevance}.  As can be seen, the weight values $\omega_{r_1}, \omega_{r_2}, \ldots \omega_{r_{|\Lambda|}}$ are arbitrarily chosen, and a mapping function $w_R(.)$  is defined to assign weights labels such that higher weight values are assigned to labels of greater interest to the community members.

\begin{algorithm}[t]
\caption{Relevance of $X_i$ (R)}
\label{alg:relevance}
\begin{algorithmic}[1]
\State Define weight values: $\omega_{r_1}, \omega_{r_2}, \omega_{r_3}, \ldots, \omega_{r_{|\Lambda|}} \in  {\mathbb R}$
\State Define a weight function $w_{R}(.)$ to assign elements in the label set $l_i \in \Lambda$ as $w_{R}(l_i) = \omega_{r_i}$
\State Compute the relevance of $X_i$ as average weighted sum:
$$R(X_i)= (\sum_{(\vec{x}_{ij}, l_{ij}) \in X_i} w_{R}(l_{ij}) )/ (\sum_{k=1}^{|\Lambda|} \omega_{r_k} )$$
\end{algorithmic}
\end{algorithm}
For each sample $x$, the corresponding label is evaluated using the mapping function $w_R(.)$ to obtain the weight value as the sample score. The relevance score of $X_i$, denoted by $R(X_i)$, is calculated as the average weight sum of the scores.

\subsubsection{Utility}
Next, we provide the sequence of steps required for calculating the utility variable of a set of indicators, in procedure~\ref{alg:utility}.  In this procedure we note that the utility of an indicator is determined by the sum of the utility weights of the samples. The weights  $\omega_{t_1}, \omega_{t_2} \ldots \omega_{t_d}$ and weight function $w_T(.)$ are defined by the application.
 
\begin{algorithm}[t]
\caption{Utility of $X_i$ (U)}
\label{alg:utility}
\begin{algorithmic}[1]
\State Define utility types $t_1, t_2, \ldots , t_d \in {\mathbb R}$ 
\State Define weight values $\omega_{t_1}, \omega_{t_2} \ldots \omega_{t_d}$, where each weight value corresponds to a utility type.
\State Define a weight function $w_T(.)$ s.t. $\ell \in \Lambda$ maps to a utility weight, i.e. $w_T(\ell) = \omega_{t_m}$ where $m= \left\{1,2, \ldots, d \right\}$.
\For{$x \in X_i$}
	\State Compute a weight of $x=(\vec{x},l^\prime)$, using $w_T(l^\prime)=\omega_{t_{l^\prime}}$
\EndFor
\State Compute the utility score of $X_i$ as the average sum of the sample weights: $U(X_i) = \frac{1}{k} \sum_{j=1}^{k} \omega_{t_j},$ where $t_j$ is the corresponding label type of sample $x_{ij} \in X_i$
\end{algorithmic}
\end{algorithm}

\subsubsection{Uniqueness}
Another metric of \qoi is their uniqueness, where highly unique indicators are considered more valuable than common indicators. In procedure~\ref{alg:uniqueness} we outline the steps used for calculating the uniqueness of a set of indicators.

\begin{algorithm}[t]
\caption{Uniqueness of $X_i$ (N)}
\label{alg:uniqueness}
\begin{algorithmic}[1] 
\State Consider the set $Z$ which is initially empty, i.e. $Z=\phi$
\State Build the set $Z$ by considering unique samples from the sets $X_1, X_2, ... X_n$
\For{$i=1,2, \ldots n $}
\For{$j=1,2, \dots k$}
  \If{$x_{ij} \notin Z$}~{add $x_{ij}$ to $Z$.}
\EndIf
\EndFor
\EndFor
\State Compute the uniqueness score for $X_i$ as follows
\begin{align}
s_n(x_{ij})=
\begin{cases}
  1 & x_{ij} \in Z \backslash \left\{X_i \right\} \\
  0 & otherwise
\end{cases}
\end{align}
\State Compute the uniqueness score (N) of $X_i$ by accumulating the scores of the elements:
$N(X_i)= \frac{1}{k} \sum_{j=1}^{k} s_n(x_{ij})$

\end{algorithmic}
\end{algorithm}

In this procedure, we assume that samples can be uniquely identified (e.g. using hashes). In set notation, we can say that an  element $x_{ij} \in X_i$ is unique if it is not an element of other sample sets, i.e. $x_{ij} \notin \bigcup X \backslash\left\{X_i \right\}$.

\subsubsection{Quality of Indicator (QoI)}
\qoi is a comprehensive measure of the various notions of quality defined earlier. In particular,  \qoi for $X_i$ is the average weighted sum of the four components: correctness (C), relevance (R), utility (U) and uniqueness (N), as shown in procedure~\ref{alg:qoi}. The weights assigned for individual metrics are application- and community member-specific. 
 \begin{algorithm}[t]
\caption{QoI of $X_i$ (QoI)}
\label{alg:qoi}
\begin{algorithmic}[1] 
\State Define normalized weights for the components: $\omega_C, \omega_R, \omega_U$, and $\omega_N$.
\State Calculate the quality of indicator (QoI) as the weighted sum of the components:
\noindent$QoI(X_i) = \omega_C C(X_i) + \omega_R R(X_i) + \omega_U U(X_i) + \omega_N N(X_i)$
\end{algorithmic}
\end{algorithm}

\section{Evaluation and Findings}\label{sec:results}
In this section, we evaluate the scoring method for contribution based on quality of indicators, and highlight how it addresses the free-riding problem in information sharing in a unique way. We start by analyzing the dataset that we obtained from AV vendors about their sample labeling, then we utilize this dataset and apply \qoi-based and volume-based scoring methods to compare between the vendors.

\subsection{Dataset Characteristics}

To highlight \qoi as a new notion of evaluating contribution in information sharing for threat intelligence, we compare the difference between quality-based and volume-based scoring methods for the contribution of AV vendors. To this end, our dataset enumerates AV vendors who submitted their artifacts of malware samples, including labels, to VirusTotal during the period of our data collection from mid 2011 to mid 2013~\cite{Mohaisen2014}. A key goal of the evaluation is to demonstrate the deficiency in the use of volume-based scores, since one vendor can achieve a high rating by submitting a large number of artifacts about malware samples of low quality. As discussed previously, this could happen because of several reasons: the submitted artifacts about some malware sample are incorrect, the sample family is uninteresting, or that the kind of information submitted about the samples are not helpful in identifying or detecting them. 

Table~\ref{tab:malfam} depicts the malware families used in this study, their sample size, and the corresponding brief description of each family type. All scans are carried out on those malware samples around May 2013 timeframe. The dataset provides a diverse representation of families, which nicely facilitate our study. As can be seen, Avzhan and Darkness are the most popular DDoS malware being submitted. On the other hand, ShadyRAT is the most popular targeted malware with the largest sample size in its category (represents about 43\% of the targeted samples and 24\% of the total samples), while Zeus has the largest sample size for Trojan malware, roughly about 77\% of the Trojan samples and 16\% of the total samples. Furthermore, we observe that more than half of the samples are DDoS (54\%), 21\% are Trojans, and 25\% are targeted malware.This breakdown provides an insight about the threat landscape and the frequency in which these types of malware appear in the wild. Oftentimes AV vendors harvest malware samples by utilizing deployment of Internet sensors for packet capturing, or using isolated environments such as honeypots and virtualization tools for behavioral analysis. The increased number of samples for DDoS is justified by the need for vast deployment for scaling up the number of infected hosts to launching attacks. On the other hand, targeted malware are less common in the wild because they are deployed in limited number of hosts, and are typically designed with covertness in mind. In the rest of this analysis, the identity of the vendors is anonymized. 


\begin{table}
\begin{center}
\caption{Malware families used in the study. DDoS stands for distributed denial of service. Also, Ddoser is known as BlackEnergy while Darkness is known as Optima. Dataset and description are from \cite{Mohaisen2014}.}\label{tab:malfam}
{\small
\begin{tabular}{|p{2cm}|p{.7cm}|p{4.4cm}|}\hline
Malware family	& \# &	Description	\\ \hline
Avzhan & 3458 & Commercial DDoS bot \\ \hline
Darkness & 1878 & Commercial DDoS bot \\ \hline
Ddoser & 502 & Commercial DDoS bot \\ \hline
jkddos & 333 & Comercial DDoS bot \\ \hline
N0ise & 431 & Commerical DDoS bot \\ \hline
ShadyRAT & 1287 & targeted gov and corps \\ \hline
DNSCalc& 403 & targeted US defense companies \\ \hline
Lurid & 399 & initially targeted NGOs \\ \hline
Getkys & 953 & targets medical sector \\ \hline 
ZeroAccess & 568 & Rootkit, monetized by click-fraud \\ \hline 
Zeus & 1975 & Banking, targets credentials \\ \hline
\end{tabular}
}
\end{center}
\end{table}

\subsection{Results}
In this section, we introduce the results and finding by performing an evaluation of the various \qoi metrics over our evaluation dataset (we evaluate all but uniqueness, since it is trivial to assess). 
First, we note that while there are more samples gathered for DDoS-type malware in comparison with others, the threat-intelligence community often gives more weight to identify malware or incidents that are less observable, which present a level of sophistication. Thus, for our evaluation, we consider trojan and targeted malware more relevant than DDoS, from the point of view of community members consuming the shared information. 

\begin{figure*}[htbp]
\begin{center}
\includegraphics[width=0.95\textwidth, height = 0.18 \textheight]{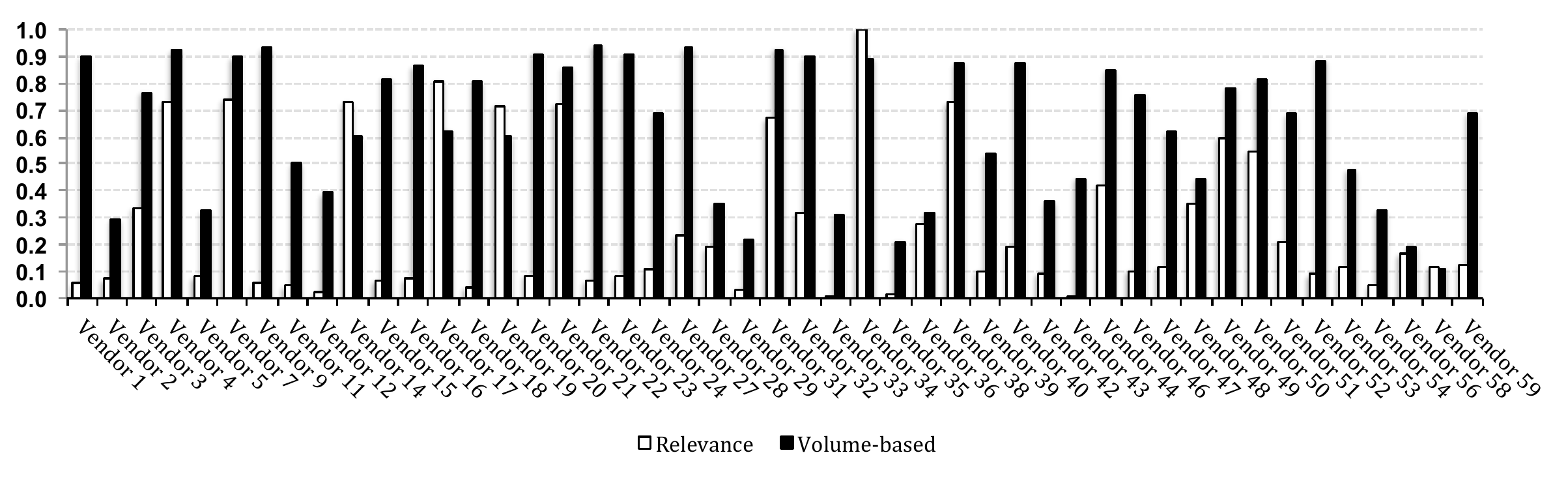}
\caption{Comparison between relevance-based and volume-based scoring}
\label{fig:relevance}
\end{center}
\end{figure*}

\BfPara{Relevance} Figure~\ref{fig:relevance} presents the normalized scores for the relevance of indicators for each vendor. In assessing relevance, we give more weight to targeted malware and Trojans over DDoS samples of each vendor (community member). Specifically, the weights are $\omega_{targeted}=5$, $\omega_{troj}=3$, and $\omega_{ddos}=1$m and ``0'' otherwise. As shown, in the relevance-based assessment a higher score is given to vendors who contribute more targeted and Trojan samples, and vendors who only contribute DDoS samples are greatly de-emphasized. 
We observe two distinct behaviors. First certain contributors who have high volume-based score tend to have a very low (close to ``0'') score when using the \qoi metric of relevance for their evaluation. In particular, with the two relevant and one less relevant family types of interest identified, such community members tend have more unidentified (irrelevant; i.e., individual score of ``0'') malware samples and families (e.g., {\tt vendor 7}, {\tt vendor 21}, {\tt vendor 59}, etc.). On the other hand, certain contributors (although smaller in numbers), and despite having a small volume-based contribution, tend to have higher relevance score, thanks to having the very relevant family identified in their shared indicator labels (e.g., {\tt vendor 10 }, {\tt vendor 16}, {\tt vendor 27}, etc.).

\begin{figure*}[htbp]
\begin{center}
\includegraphics[width=0.95\textwidth, height = 0.18 \textheight]{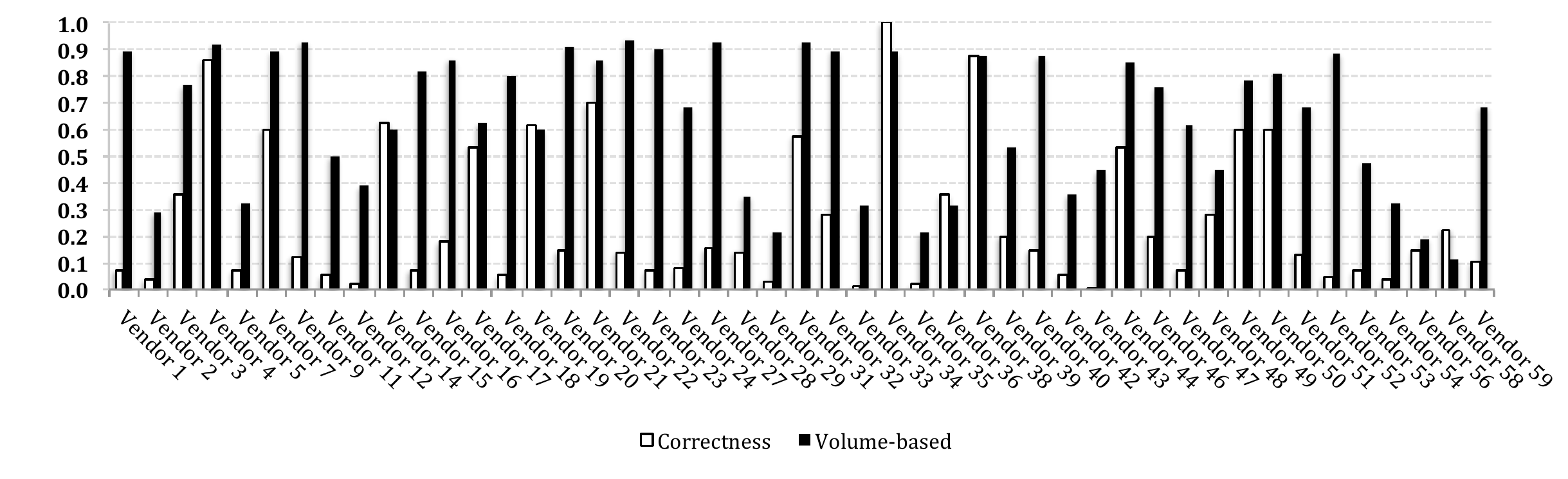}
\caption{Comparison between correctness-based and volume-based scoring of AV indicators}
\label{fig:correct}
\end{center}
\end{figure*}

\BfPara{Correctness} An assessment of the correctness of the AV indicators is depicted in Figure~\ref{fig:correct}. As can be seen, vendor 4, vendor 27, and vendor 30, outperformed other vendors in this metric with a score in the 80s up to top 90s percentile range, highlighting a quality of the labels provided by those vendors corresponding to the actually learned labels of the samples provided by them. On the other hand, the majority of the rest of vendors tend to have a gap in the score between the volume-based and the correctness-based contribution measures, where the correctness-based measure is significantly lower. There are various reasons for why some vendors tend to score low for correctness despite their large (volume-based) contribution. This could be caused by them labeling samples under unknown names, mislabeling  to other family types due to similarities between families, or assigning generic labels like ``{\tt trojan}'', ``{\tt virus}'', ``{\tt unclassified}'' among other misleading labels. Examining the correctness of AV indicators also leads to a more subtle discussion about their utility. Looking closer into the label generated by some vendors, we find out that some labels are too generic in the sense that they only describe the behavior rather than name of a known malware family type, e.g., {\tt Trojan.Win32.ServStart} vs. {\tt Avzhan}. 

\BfPara{Utility} To evaluate the utility of AV indicators, we give weights for three classes of malware labels: complete labels ($\omega_c$) are based on industrially popular name, generic labels ($\omega_g$) are based on placeholders commonly used for labeling the family such as ``{\tt generic}'', ``{\tt worm}'', ``{\tt trojan}'', ``{\tt start}'' and ``{\tt run}'', and incomplete labels, ($\omega_i$), including ``{\tt suspicious}'', ``{\tt malware}'', and ``{\tt unclassified}'', which do not hold any meaning of a class. Similar to the strategy with relevance, we assign weights of $\omega_c=5$, $\omega_g=2$, and $\omega_i=1$.

\begin{figure*}[htbp]
\begin{center}
\includegraphics[width=0.95\textwidth, height = 0.18 \textheight]{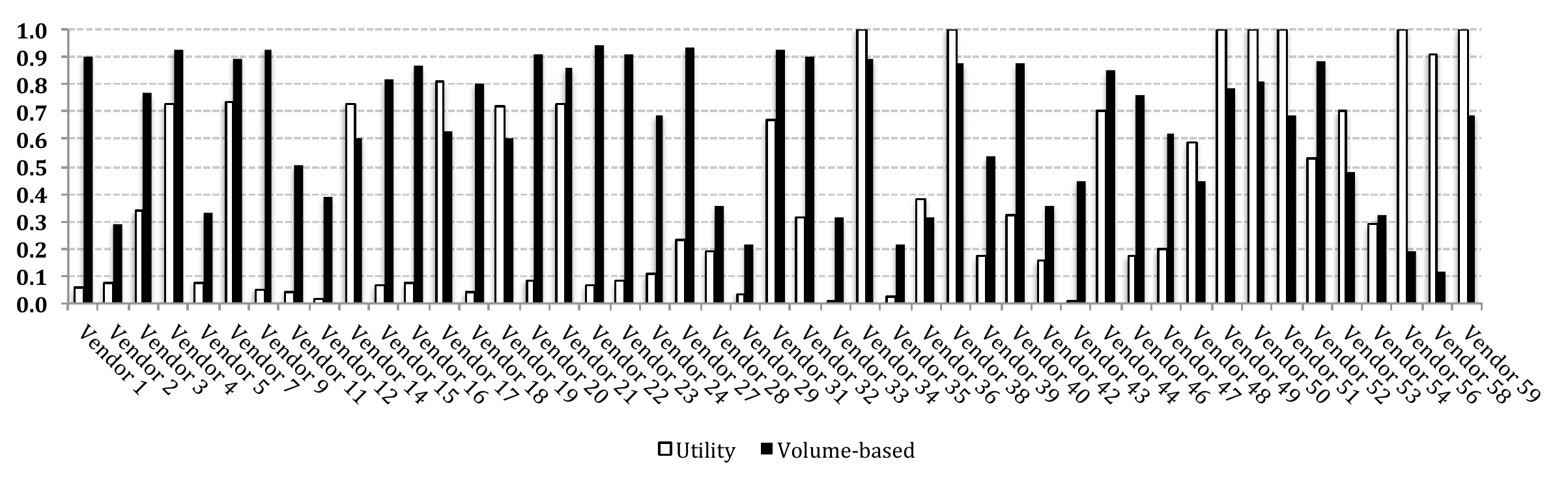}
\caption{Comparison between utility-based and volume-based scoring}
\label{fig:utility}
\end{center}
\end{figure*}

\begin{figure*}[htbp]
\begin{center}
\includegraphics[width=0.95\textwidth, height = 0.18 \textheight]{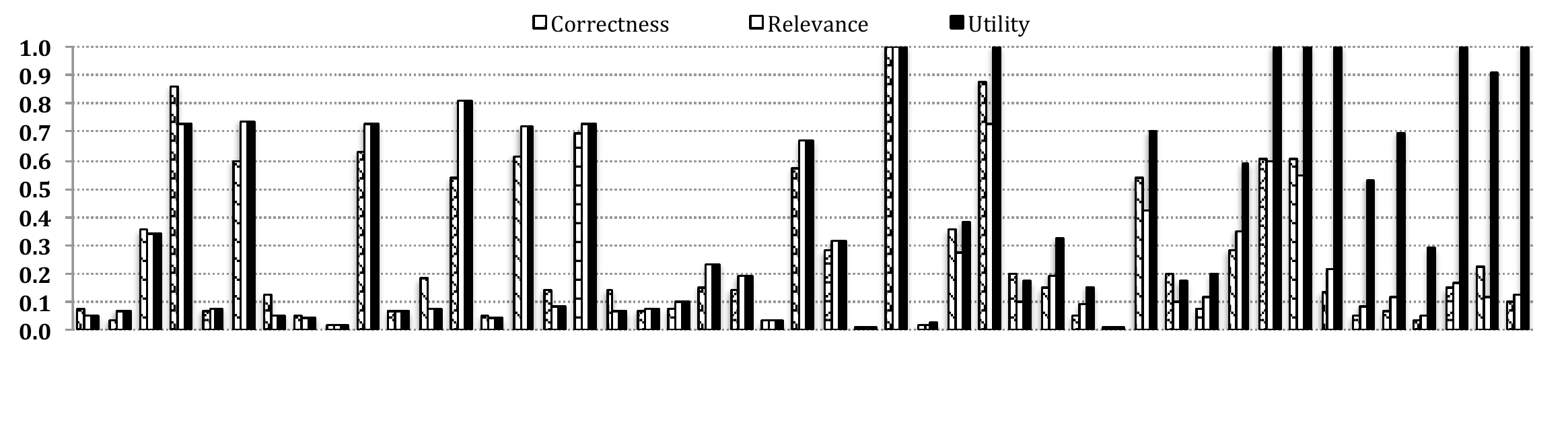}
\caption{Comparison between various quality-based metrics for AV indicator assessment}
\label{fig:metrics}
\end{center}
\end{figure*}

We plot the results of evaluating the utility of indicators in Figure~\ref{fig:utility}. We notice that vendors such as {\tt vendor 51}, {\tt vendor 53}, and {\tt vendor 59} are rated as high utility indicator providers that surpass their volume-based scores. Nevertheless, these vendors' high utility indicator is offset in Figure~\ref{fig:metrics} that includes two more metrics: correctness and relevance. These additional metrics show that these vendors achieve insignificant correctness and relevance.  Figure~\ref{fig:metrics}  captures  the importance of displaying the three scores in a single plot to allow direct comparison of the
various quality-based metrics. We notice that there is a clear correlation between the correctness and relevance scores. This is because we only compute relevance scores for correctly submitted samples, where incorrect labels are zeroed in the relevance score.

\begin{figure*}[htbp]
\begin{center}
\includegraphics[width=0.95\textwidth, height = 0.18 \textheight]{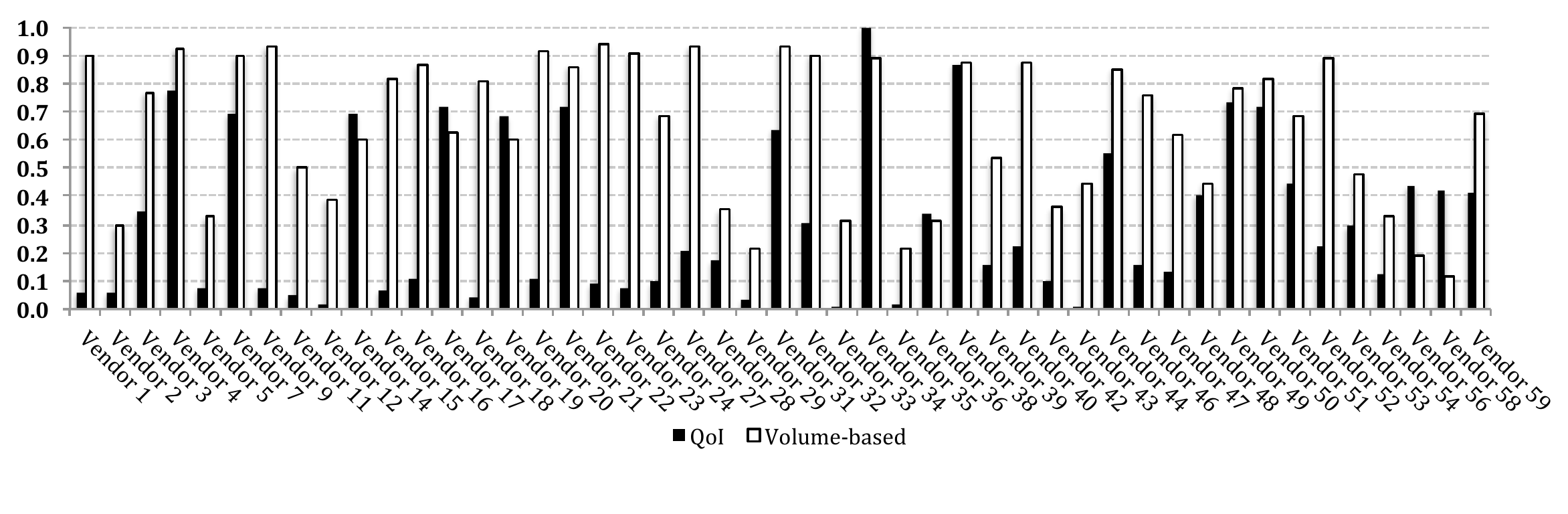}
\caption{Comparison between QoI-based and volume-based scoring}
\label{fig:qoi}
\end{center}
\end{figure*}

\BfPara{Aggregated \qoi score} As described earlier, we aggregate a single \qoi score for each vendor based on the weighted sum of the various \qoi metrics. In Figure~\ref{fig:qoi}, we present a comparative bar-plot between \qoi-based and volume-based scoring for assessment of contribution by the AV vendors. As can be seen, many vendors, such as {\tt vendor 14}, {\tt vendor 57}, and {\tt vendor 58} which received high \qoi scores are rated with lower scores in their volume-based rating. In particular, they received from 15-75 percent lower rating in their volume-based scores.  On the other hand, vendors like {\tt vendor 11} and {\tt vendor 18}, and {\tt vendor 20}, which tended to provide a very high volume-based indicators, have very small (close to zero) \qoi score, highlighting their potential as free-riding candidates.

\section{Related Work}\label{sec:related}

The free-riding problem in threat intelligence sharing is not addressed before, nor measures of quality broadly defined, or closely identified for assessing contribution by community members. However, the problem of free-riding in general is not new and has been a topic of discussion in the peer-to-peer (P2P) systems community. 

Going early back in the literature, Adar and Huberman~\cite{adar2000} were first to spot the problem in P2P systems as they noticed the existence of a large fraction of users who do not share useful content in the file sharing network Gnutella.  A few years later, Feldman et al.~\cite{Feldman2004} characterized the problem of free-riding in peer-to-peer systems and proposed potential directions for research. In response, a stream of papers were published on the topic, most notably, the work of Locher et al.~\cite{locher2006}, who developed a free-riding client as a proof-of-concept and demonstrated how entire files can be downloaded in the BitTorrent network without providing any content to the peers. While Locher et al. have concentrated on analyzing the feasibility of free-riding attacks, other papers~\cite{varian2004system, hughes2005} are more focused on analyzing their root-cause and impact on the overall system utility.

In light of operationalizing the functions of threat intelligence, various information sharing standards were proposed including those developed by IETF (\url{https://www.ietf.org/}), MITRE (\url{https://www.mitre.org/}), and NIST (\url{https://www.nist.gov/}).  Industry leaders have picked up on these standards and developed application program interfaces (APIs) to facilitate delivery and retrieval of raw, processed, and structure and intelligence data, such as ThreatExchange~\cite{ThreatExchange} by Facebook and IntelGraph by Verisign. However, sharing standards have shown to exhibit privacy violations including leaking PII fields, as demonstrated by~\cite{Mohaisen2016Rethink}, potentially encouraging the act of free-riding. 


Of relevance to the notion of quality of indicators in threat intelligent systems is malware attribution. Malware attribution have been widely employed in the literature for training algorithms and techniques of malware classification and labeling~\cite{rossow2012}, and understanding the utility of attributes as detector patterns of malware samples has been an important subject matter. Bailey et al.~\cite{bailey2007automated} were one of the early folks to characterize malware in terms of system state changes (e.g. registry changes, files created) and investigated the problem of  behavior-based clustering as a method for classifying and analyzing Internet malware.

More focused on the labeling problem, Canto et al.~\cite{canto2008} analyzed the quality of labeling of malware samples for a couple of vendors and pointed out their labeling inconsistencies. In the same vein, Perdisci~\cite{vamo2012} analyzed the shortcomings of malware labeling of various AV vendors by constructing a graph from the labels and measuring the distance between them. On the other hand, Mohaisen and Alrawi~\cite{Mohaisen2014, Mohaisen2015} quantified the inconsistencies in labeling against a reference dataset collected from thousands of samples of various types which were manually vetted by analysts. In their study, the authors evaluated the detection rate, correctness, and consistency of labeling of AV scanners.

\section{Conclusion}\label{sec:conclusion}

In this paper, we have the first look at the notion of the quality of indicators (\qoi) for understanding the contribution of community members in information sharing paradigms. Unlike other peer-to-peer systems in which the volume of contribution (bandwidth, size of files, etc.) is a good indicator of contribution, we argue that the special nature of security applications calls for more elaborate notion of contribution. As such, we define multiple metrics for assessing contribution, including correctness, utility, and relevance of indicators. As compared to volume-based measures for contribution, and thus free-riding, our metrics are more robust, contextual, and reasonably quantify the actual contribution of individuals. By verifying our metrics on a real-world data of antivirus scans we unveil that contribution measured by volume is not always consistent with those quality measures, and that \qoi as notion is capable of capturing forms of contribution beyond free-riding.  

\balance

\end{document}